\documentclass[aps,prd,twocolumn,superscriptaddress,amsfont,graphicx,nofootinbib,preprintnumbers]{revtex4}%
\pdfoutput=1
\usepackage{color,graphicx}
\usepackage{ifpdf}
\usepackage{amsmath}
\usepackage{bm}
\usepackage{color}
\usepackage[english]{babel}
\usepackage{graphicx}%
\usepackage{amsfonts}%
\usepackage{amssymb}
\usepackage{braket}
\usepackage{hyperref}

\definecolor{nicered}{rgb}{0.7,0.1,0.1}
\definecolor{nicegreen}{rgb}{0.1,0.5,0.1}
\hypersetup{colorlinks,citecolor= nicegreen,linkcolor= nicered}

\newcommand{\cH}{\mathcal H}

\newcommand{\be}{\begin{equation}}
\newcommand{\ee}{\end{equation}}
\newcommand{\bea}{\begin{eqnarrat}}
\newcommand{\eea}{\end{eqnarray}}

\definecolor{Red}{rgb}{1.,0.,0.}

\newcommand{\SM}{{\rm SM}}
\newcommand{\NP}{{\rm NP}}
\newcommand{\ntree}{{\rm peng}}

\begin{document}

\preprint{CERN-PH-TH/2011-301}

\title{Implications of the LHCb Evidence for Charm CP Violation}

\author{Gino Isidori}
\affiliation{CERN, Theory Division, CH1211 Geneva 23, Switzerland}
\affiliation{INFN, Laboratori Nazionali di Frascati, Via E. Fermi 40, 00044 Frascati, Italy}

\author{Jernej F.\ Kamenik} 
\affiliation{J. Stefan Institute, Jamova 39, P. O. Box 3000, 1001 Ljubljana, Slovenia}
\affiliation{Department of Physics, University of Ljubljana, Jadranska 19, 1000 Ljubljana, Slovenia}

\author{Zoltan Ligeti}
\affiliation{Ernest Orlando Lawrence Berkeley National Laboratory,
University of California, Berkeley, CA 94720}

\author{Gilad Perez}
\affiliation{CERN, Theory Division, CH1211 Geneva 23, Switzerland}
\affiliation{\mbox{Department of Particle Physics and Astrophysics, 
Weizmann Institute of Science, Rehovot 76100, Israel}}

\date{\today}
\begin{abstract}

The LHCb collaboration recently announced preliminary evidence for CP violation
in $D$ meson decays. We discuss this result in the context of the standard model
(SM), as well as its extensions. In the absence of reliable methods to evaluate
the hadronic matrix elements involved, we can only estimate qualitatively
the magnitude of the non-SM tree level operators required to generate the
observed central value. In the context of an effective theory, we list the
operators that can give rise to the measured CP violation and investigate
constraints on them from other processes.

\end{abstract}

\maketitle

\section{Introduction}

Recently the LHCb collaboration reported a $3.5\sigma$ evidence for a non-zero
value of the difference between the time-integrated CP asymmetries in the
decays $D^0 \to K^+K^-$ and $D^0 \to \pi^+\pi^-$~\cite{lhcb},
\begin{equation}\label{LHCbnews}
\Delta a_{CP} \equiv a_{K^+ K^-} - a_{\pi^+ \pi^-} 
 = -(0.82\pm 0.21\pm0.11)\%\,.
\end{equation}
The time-integrated CP asymmetry for a final CP eigenstate, $f$, is
defined as
\begin{equation}
a_f \equiv \frac{\Gamma(D^0\to f)-\Gamma(\bar D^0\to f)}
  {\Gamma(D^0\to f)+\Gamma(\bar D^0\to f)}\,.
\end{equation}
Combined with previous measurements of these CP
asymmetries~\cite{Aaltonen:2011se, Staric:2008rx, Aubert:2007if, HFAG}, the
world average is
\begin{equation}
\Delta a_{CP} = -(0.65\pm 0.18)\%\,.
\label{eq:acpExp}
\end{equation}

Following~\cite{Grossman:2006jg} we write the singly-Cabibbo-suppressed $D^0\
(\bar D^0)$ decay amplitudes $A_f\ (\bar A_f)$ to CP eigenstates, $f$, as
\begin{subequations}
\begin{eqnarray}
A_f &=& A^T_f\, e^{i \phi_f^T} \big[1+r_f\, e^{i(\delta_f + \phi_f)}\big]\,, \\
\bar A_f &=& \eta_{CP}\, A^T_f\, e^{-i \phi_f^T} 
  \big[1+r_f\, e^{i(\delta_f - \phi_f)}\big]\,, 
\end{eqnarray}
\end{subequations}
where $ \eta_{CP}=\pm1$ is the CP eigenvalue of $f$, the dominant
singly-Cabibbo-suppressed ``tree" amplitude is denoted $A^T_f\, e^{\pm i
\phi^T_f}$, and $r_f$ parameterizes the relative magnitude of all the subleading
amplitudes (often called ``penguin" amplitudes), which have different strong
($\delta_f$) and  weak ($\phi_f$) phases. 

In the following we focus on the $\pi^+ \pi^-$ and $K^+ K^-$ final states. In
general, $a_f$ can be written as a sum of CP asymmetries in decay, mixing, and
interference between decay with and without mixing.  Mixing effects are
suppressed by the $D^0-\bar D^0$ mixing parameters, and, being universal, tend
to cancel in the difference  between $K^+ K^-$ and $\pi^+ \pi^-$ final
states~\cite{Grossman:2006jg}. Taking into account the different time-dependence
of the acceptances in the two modes, LHCb quotes~\cite{lhcb} for the
interpretation of Eq.~(\ref{LHCbnews}),
\be
a_{K^+ K^-} - a_{\pi^+\pi^-} \approx a^{\rm dir}_K - a^{\rm dir}_\pi
  + (0.10\pm0.01)\,a_{\rm ind}\,.
\ee
Thus, {because of the experimental constraints on the mixing parameters [see
Eq.~(\ref{eq:expth})]}, a large  $\Delta a_{CP}$ can be generated only by  the
direct CP violating terms,
\begin{equation}
a_f^{\rm dir} = -\frac{2 {r_f} \sin \delta_f \sin \phi_f}
  {1+2 r_f \cos \delta_f \cos \phi_f + r_f^2}\,,
\label{eq:adf}
\end{equation}
and we use the $f=K,\pi$ shorthand for $K^+K^-$ and $\pi^+\pi^-$.

\section{General considerations and SM prediction}

Independent of the underlying physics, a necessary condition for non-vanishing
$a_f^{\rm dir}$ is to have at least two amplitudes with different strong and
weak phases contribute to the final state $f$. In the isospin symmetry limit,
the condition on the strong phases implies that different isospin amplitudes
have to contribute. Since the leading (singly-Cabibbo-suppressed) terms in the
standard model (SM) effective Hamiltonian, defined below, have both $\Delta
I=1/2$ and $\Delta I=3/2$ components, the subleading operators with a different
weak phase may have a single isospin component.  As far as amplitudes with a
different weak phase are concerned, in the SM, as well as within its MFV
expansions~\cite{MFV, GMFV}, they are suppressed by $\xi \equiv |V_{cb}V_{ub}| /
|V_{cs}V_{us}| \approx 0.0007$.  

The SM effective weak Hamiltonian relevant for hadronic 
singly-Cabibbo-suppressed $D$ decays, renormalized at a scale $m_c < \mu< m_b$ 
can be decomposed as 
\begin{equation}\label{SMHeff}
\mathcal H^{\rm eff}_{|\Delta c| = 1} = 
  \lambda_d\, \mathcal H^{d}_{|\Delta c| = 1}
  + \lambda_s\, \mathcal H^{s}_{|\Delta c| = 1}
  + \lambda_b\,  \mathcal H^{\ntree}_{|\Delta c| = 1}\,,
\end{equation}
where $\lambda_q = V_{cq}^*V_{uq}$, and
\begin{eqnarray}\label{eq:Q12} 
\mathcal H^{q}_{|\Delta c| = 1} &=& \frac{G_F}{\sqrt 2}
  \sum_{i=1,2} C^q_i Q_i^s + {\rm H.c.}\,, \qquad q=s,d, \nonumber \\
Q^q_1 &=& (\bar u  q)_{V-A}\, (\bar q c)_{V-A}\,, \nonumber \\  
Q^q_2 &=& (\bar u_\alpha  q_\beta)_{V-A}\, (\bar q_\beta  c_\alpha)_{V-A}\,,
\end{eqnarray}
and $\alpha,\beta$ are color indices.
The first two terms in Eq.~(\ref{SMHeff}) have ${\cal O}(1)$ Wilson coefficients
in the SM. On the contrary, the so-called penguin operators in
$\mathcal H^{\ntree}_{|\Delta c| = 1}$ have tiny Wilson coefficients at scales $m_c <
\mu< m_b$ (see Refs.~[\onlinecite{Golden:1989qx}, \onlinecite{Grossman:2006jg}]
for the list of relevant operators and Wilson coefficients).

Let us first consider the $D \to K^+K^-$ amplitude. In the SM, it is convenient
to use CKM unitarity, $\lambda_d + \lambda_s + \lambda_b=0$, to eliminate the
$\lambda_d$ term, and obtain $A_{K} = \lambda_s (A^s_{K} - A^d_{K}) + \lambda_b
(A_{K}^b - A_{K}^d)$. For $D \to \pi^+\pi^-$, it is convenient to eliminate
$\lambda_s$ to obtain $A_\pi = \lambda_d (A^d_\pi - A^s_\pi) + \lambda_b
(A_\pi^b - A_\pi^s)$.  This way, the first terms are singly-Cabibbo-suppressed,
while the second terms are both CKM suppressed and have either  vanishing
tree-level matrix elements or tiny Wilson coefficients.  The magnitudes of these
subleading amplitudes are controlled by  the CKM ratio $\xi=
|\lambda_b/\lambda_s| \simeq |\lambda_b/\lambda_d| \approx 0.0007$ and the
ratio  of hadronic amplitudes.  We define
\be\label{RKpidef}
R^{\SM}_K =  \frac{ A^b_K - A^d_K }{ A_K^s - A_K^d }\,, \qquad 
R^{\SM}_\pi = \frac{ A^b_\pi - A^s_\pi }{ A_\pi^d - A_\pi^s }\,.
\ee
Since $\mathrm{arg}(\lambda_b/\lambda_s) \approx
-\mathrm{arg}(\lambda_b/\lambda_d) \approx 70^\circ$, we can set
$|\sin(\phi^{\SM}_f)| \approx 1$ in both channels, and neglect the interference
term in the denominator of Eq.~(\ref{eq:adf}). 

In the $m_c \gg \Lambda_{\rm QCD}$ limit, one could analyze these decay
amplitudes model independently.  Given the valence-quark structure of the
$K^+K^-$ final state, a penguin contraction is required for operators  of the
type $c \to ud\bar d$ or $ub\bar b$ to yield a non-vanishing $D \to K^+K^-$
matrix element. This is why $R^{\SM}_K$ is expected to be substantially smaller
than one.  A na\"ive estimate in perturbation theory yields $|A^d_K/A^s_K| \sim
\alpha_s(m_c)/\pi\sim 0.1$  (and $|A^b | \lesssim |A^d |$). However, since the charm
scale is not far from $\Lambda_{\rm QCD}$, non-perturbative enhancements leading
to substantially larger values cannot be excluded~\cite{Golden:1989qx}.  The
same holds for the ratio $R^{\SM}_\pi$ defined in Eq.~(\ref{RKpidef}). 

To provide a semi-quantitative estimate of $R^{\SM}_{K,\pi}$ beyond perturbation
theory, we note that  penguin-type contractions are absent in the
Cabibbo-allowed $c\to u s \bar d$ Hamiltonian, contributing to $D\to K^+ \pi^-$.
In the absence of penguin contractions, $D\to K^+K^-$ and  $D\to \pi^-\pi^+$
amplitudes have identical topologies to $D\to K^+ \pi^-$, but for appropriate
$s\leftrightarrow d$ exchanges of the valence quarks. The data imply  $|A_{KK}|
\approx 1.3\, |\lambda_s  A_{K\pi}|$ and $A_{\pi\pi} \approx 0.7\, |\lambda_s
A_{K\pi}|$.  These results are compatible with the amount of $SU(3)$ breaking
expected in the tree-level amplitudes and show no evidence for  anomalously
large penguin-type contractions competing with the tree-level amplitudes.
Further evidence that tree-level topologies dominate the decay rates is
obtained from the smallness of $\Gamma(D\to K^0\bar K^0)/\Gamma(D\to K^+K^-)$,
which is consistent with the vanishing $D\to K^0\bar K^0$ tree-level matrix
element of $\cH^{(s -d)}$ in the $SU(3)$ limit. However, it must be stressed
that data on the decay  rates do not allow us to exclude a substantial
enhancement of the CKM suppressed amplitudes.  The latter do not have an $s-d$
structure as the leading Hamiltonian, and, if enhanced over na\"ive estimates as
in the case of the $\Delta I =1/2$ rule in  $K\to \pi\pi$ amplitudes, may
account for $|R^{\SM}_{K,\pi}|>1$~\cite{Golden:1989qx}.

In the following we assume that $r_f \ll 1$ even in the presence of new physics
(NP), and we can expand Eq.~(\ref{eq:adf}) to first order in this parameter. We
can thus write 
\be
a_K^{\rm dir}  \approx 2 \bigg[ \xi\, \mathrm{Im} (R_K^{\SM})
  + \frac1{\sin\theta_C} \sum_i \mathrm{Im}(C_i^{\NP})\, \mathrm{Im}
  (R_{K,i}^{\NP}) \bigg] ,
\ee
and similarly in the $\pi^+\pi^-$ mode. Here  $R_{K,i}^{\NP}$ denote the ratio
of the subleading amplitudes generated by the operators $Q_i$ in the NP
Hamiltonian defined below in Eq.~(\ref{eq:HNP}), normalized to the dominant SM amplitude,
after factoring out the leading CKM dependence, $\sin\theta_C \approx
|\lambda_{s,d}|\approx 0.225$, and the NP Wilson coefficients,\footnote{Contrary to the SM case, 
   where the CKM factors are explicitly factorized, in the NP case
   we include flavor mixing terms in the $C_i^{\rm NP}$ --- see Eq.~\eqref{eq:HNP}.} $C^\NP_i$. This
implies
\be
\Delta a_{CP}  \approx (0.13\%) \mathrm{Im} (\Delta R^{\SM})
  + 8.9 \sum_i \mathrm{Im} (C_i^{\NP})\, \mathrm{Im} (\Delta R^{\NP}_{i} ),
\label{eq:acpNP}
\ee
where we defined
\be\label{DeltaRdef}
\Delta R^{\SM,\NP} = R_K^{\SM,\NP}  + R^{\SM,\NP} _\pi\,.
\ee
In the $SU(3)$ limit, $R_K^{\SM} = R_\pi^{\SM}$, and therefore $a_K^{\rm dir}
\approx - a_\pi^{\rm dir}$, which add constructively in $\Delta
a_{CP}$~\cite{Golden:1989qx, Quigg:1979ic}.

\begin{figure}[tb]
  \centering
  \includegraphics[width=0.47\textwidth]{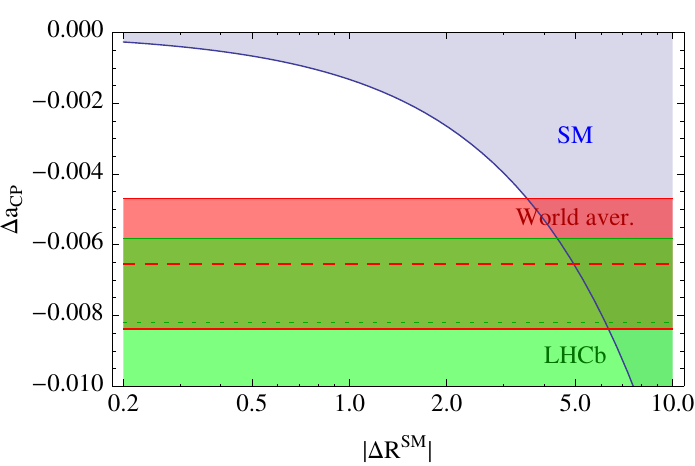}
  \caption{Comparison of the experimental $\Delta a_{CP}$ values with the SM
  reach as a function of $|\Delta R^{\SM}|$. } \label{fig:SM}
\end{figure}

Assuming the SM, the central value of the experimental result is recovered if
${\rm Im}(\Delta R^{\SM}) \approx 5$, as illustrated in Fig.~\ref{fig:SM}.  Such
an enhancement of the CKM-suppressed amplitude cannot be excluded from first
principles, but it is certainly beyond its na\"ive
expectation~\cite{Grossman:2006jg}.

Note that the applicability of $SU(3)$ flavor symmetry should be questioned,
because the $D\to K^+K^-$ and $D\to \pi^+\pi^-$ decay rates imply a large
breaking of the symmetry.  Without $SU(3)$ as a guidance, one can no longer
expect $a_K^{\rm dir} \approx - a_\pi^{\rm dir}$; in particular, the strong
phases relevant for direct CP violation in these two channels are no longer
related.  One might then expect $|a_\pi^{\rm dir}| < |a_K^{\rm dir}|$, if the
deviation from factorization is smaller in the $\pi^+\pi^-$ than in the $K^+K^-$
mode.  Therefore, it will be very interesting for the interpretation of the
results  when the CP asymmetries are measured separately with increased
precision.  Recent measurements by CDF~\cite{Aaltonen:2011se},
Belle~\cite{Staric:2008rx} and BaBar~\cite{Aubert:2007if} yield for the average
of the individual CP asymmetries (without LHCb, and dominated by
CDF~\cite{Aaltonen:2011se}) in the $\pi^+\pi^-$ and $K^+K^-$ modes
$(2.0\pm2.2)\times 10^{-3}$ and $(-2.3\pm1.7)\times 10^{-3}$~\cite{HFAG},
respectively, which does not yet allow us to draw definite conclusions [and are
included in Eq.~(\ref{eq:acpExp})].
%
%
  Another important experimental handle to decide whether the
observed signal can or cannot be accommodated in the SM would be observing or
constraining CP violation in other decay modes, corresponding to the same
quark-level transitions.  These include pseudoscalar-vector or vector-vector
final states, three-body decays, $D_s$ and $\Lambda_c$ decays.  More precise
measurements in such decays will help to decide whether the measured CP
asymmetry in Eq.~(\ref{LHCbnews}) is due to new short distance physics, or to a
large enhancement of a hadronic matrix element in one particular channel.

\section{New Physics contributions}

The size of NP effects allowed in $\Delta a_{CP}$ depends on $\mathrm{Im} (\Delta R^{\SM})$. In order to understand the scale probed by the
measurement, we parametrize the NP contributions in terms of an effective NP
scale $\Lambda_{\rm NDA}$, normalized to the Fermi scale:  $
\mathrm{Im}(C_i^{\NP})  =  \sqrt 2\, \mathrm{Im} (C_{\rm NDA}) / (\Lambda_{\rm
NDA}^2 G_F)$. The resulting sensitivity for $\Lambda_{\rm NDA}$ ($C_{\rm
NDA}$) can be written as
\be\label{eq:NP}
\mathrm{Im}(C_{\rm NDA})\,\frac{(10~\rm TeV)^2}{\Lambda_{\rm NDA}^2}
  = \frac{ {(0.61\pm 0.17)} - 0.12\, \mathrm{Im}(\Delta R^\SM)}{\mathrm{Im}(\Delta R^\NP)}. 
\ee
In other words, assuming ${\mathrm{Im}(\Delta R^\NP)}\sim1$, $|\Delta R^\SM| \ll
5 $ and $C_{\rm NDA}=1$ implies that a NP scale of $\mathcal
O(13{\rm\,TeV})$ will saturate the observed CP violation; alternatively,
setting  $\Lambda_{\rm NDA}\to 2^{1/4}/\sqrt{G_F}$ implies that $C_{\rm
NDA} \sim 7 \times10^{-4}$ is required.  As we discuss below, despite
the large scale involved, after taking into account the bounds from CP violation
in $|\Delta c|=2$ and $|\Delta s|=1$ processes, only a few NP  operators may
saturate the value in Eq.~(\ref{eq:NP}) in the limit $|\Delta R^\SM| \ll 5 $.

To discuss possible NP effects, we consider the following effective Hamiltonian
\begin{eqnarray}
\mathcal H^{\rm eff-\rm NP}_{|\Delta c| = 1} &=& \frac{G_F}{\sqrt 2} 
  \sum_{i=1,2,5,6} \sum_q  (C^q_i  Q_i^q + C^{q\prime}_i Q^{q\prime}_i) \nonumber \\
&+& \frac{G_F}{\sqrt 2} \sum_{i=7,8}  (C_i  Q_i + C'_i Q^{\prime}_i)+ {\rm H.c.}\,,
\label{eq:HNP}
\end{eqnarray}
where $q=\{d,s,b,u,c\}$, and the list of operators includes, in addition to
$Q^q_{1,2}$ given in Eq.~(\ref{eq:Q12}), 
\begin{eqnarray}
Q^q_5 &=& (\bar u  c)_{V-A}\, (\bar q  q)_{V+A}\,,\nonumber\\
Q^q_6 &=& (\bar u_\alpha c_\beta)_{V-A}\, (\bar q_\beta  q_\alpha)_{V+A}\,, \nonumber\\
Q_{7} &=& - \frac{e}{8\pi^2}\, m_c\, \bar u \sigma_{\mu\nu} 
  (1+\gamma_5) F^{\mu\nu}\, c\,, \nonumber\\
Q_{8} &=& - \frac{g_s}{8\pi^2}\, m_c\, \bar u \sigma_{\mu\nu} 
  (1+\gamma_5) T^a G_a^{\mu\nu} c\,, 
\end{eqnarray}
and another set, $Q^{(q)\prime}_i$, obtained from $Q^{(q)}_i$ via the
replacements $A \leftrightarrow -A$ and $\gamma_5 \leftrightarrow -\gamma_5$.
This is the most general dimension-six effective Hamiltonian relevant for $D\to
K^+K^-,\, \pi^+\pi^-$ decays, after integrating out heavy degrees of freedom 
around or above the electroweak scale.

\subsection{\boldmath Bounds on NP effects from $D^0 -\bar D^0$ mixing}

Charm mixing arises from $|\Delta c| = 2$ interactions that generate
off-diagonal terms in the mass matrix for $D^0$ and $\bar D^0$ mesons. The $D^0
-\bar D^0$ transition amplitudes are defined as
\begin{equation}
\bra{D^0} \mathcal H \ket{\bar D^0} = M_{12} - \frac{i}{2} \Gamma_{12}\,.
\end{equation}
The three physical quantities related to the mixing can be defined as
\begin{equation}
y_{12} \equiv \frac{|\Gamma_{12}|}{\Gamma}\,, \quad 
x_{12} \equiv 2 \frac{|M_{12}|}{\Gamma}\,, \quad 
\phi_{12} \equiv \mathrm{arg} \bigg(\frac{M_{12}}{\Gamma_{12}}\bigg)\,.
\end{equation}
HFAG has performed a fit to these theoretical quantities, even allowing for CP
violation in decays, and obtained the following 95\% C.L.\ regions~\cite{HFAG}
\begin{align}
x_{12} &\in [0.25,\, 0.99]\,\%\,, \nonumber\\ 
y_{12} &\in [0.59,\, 0.99]\,\%\,, \nonumber\\
\phi_{12} &\in [-7.1^\circ,\, 15.8^\circ]\,.
\label{eq:expth}
\end{align}
We cannot reliably estimate the SM contributions to these quantities from first
principles, and thus simply require the NP contributions to at most saturate
the above experimental bounds on $x_{12},\ y_{12}$, and $\phi_{12}$.

\begin{figure}[t]
  \centering
  \includegraphics[width=0.2\textwidth]{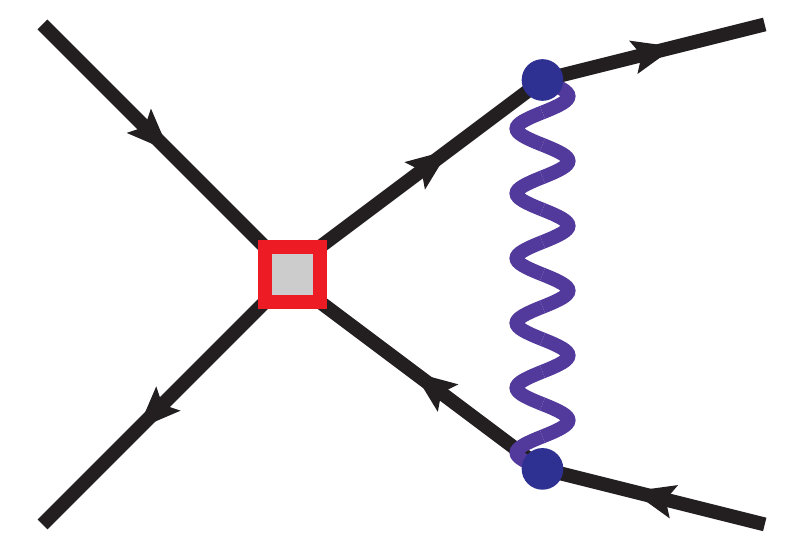}
  \caption{Contribution of $\mathcal H^{\rm
eff-NP}_{|\Delta c| = 1}$ (red square) to $|\Delta c|=2$ and $|\Delta s|=1$ operators via a $W$ (blue wavy line) loop. 
  \label{fig:dressing}}
\end{figure}

The NP operators present in $\mathcal H^{\rm eff-NP}_{|\Delta c| = 1}$ may
affect  $D^0$--$\bar D^0$ mixing parameters at the second order in the NP coupling,
$T\big\{\mathcal H^{\rm eff-NP}_{|\Delta c| = 1}(x)\,  \mathcal H^{\rm
eff-NP}_{|\Delta c| = 1} (0)\big\}$.  {Such a contribution, which formally
corresponds to a quadratically divergent one loop  diagram, is highly UV
sensitive. If we assume a fully general structure for our effective theory,
where operators are of NDA strength, then the scaling in Eq.~(\ref{eq:NP}) would
imply much too large contributions to $D-\bar D$ mixing and CP violation (see,
e.g.,~\cite{Isidori:2010kg}).  This could be a major constraint for many
SM extensions. However, being a genuine UV effect, it is also 
highly model dependent.  On the other hand, assuming that $\mathcal H^{\rm
eff-NP}_{|\Delta c| = 1}$ is generated above the electroweak scale and the UV
completion of the theory  cures the above mentioned problem, we can   derive
(model-independent) bounds on the coefficients of $\mathcal H^{\rm
eff-NP}_{|\Delta c| = 1}$ from the effective $|\Delta c|=2$ operators generated at low energies, by considering its time ordered product with the SM charged-current interactions, $T\big\{\mathcal H^{\rm eff-NP}_{|\Delta c| = 1}(x)\,  \mathcal
H^{\rm SM}_{|\Delta c| = 1} (0)\big\}$ (schematically depicted Fig.~\ref{fig:dressing}). }

The effective Hamiltonian thus obtained integrating out all the heavy fields is
\begin{equation}
\mathcal H^{\rm eff}_{|\Delta c|=2} = \frac{G_F}{\sqrt 2} 
  \left( \sum_{i=1}^5 C^{cu}_i Q_i^{cu} + \sum_{i=1}^3 C^{cu\prime}_i 
  Q_i^{cu\prime}  \right) ,
\label{eq:HdC=2}
\end{equation}
where
\begin{eqnarray}
Q_1^{cu} &=& (\bar u c)_{V-A} \, (\bar u c)_{V-A}\,, \nonumber\\
Q_2^{cu} &=& (\bar u c)_{S-P} \, (\bar u c)_{S-P}\,, \nonumber\\
Q_3^{cu} &=& (\bar u_\alpha c_\beta)_{S-P} \, (\bar u_\beta c_\alpha)_{S-P}\,, \nonumber\\
Q_4^{cu} &=& (\bar u c)_{S-P} \, (\bar u c)_{S+P}\,,\nonumber\\
 Q_5^{cu} &=& (\bar u_\alpha c_\beta)_{S-P} \, (\bar u_\beta c_\alpha)_{S+P}\,,
\end{eqnarray}
and, as before, the $Q_{1,2,3}^{cu\prime}$ operators are obtained from
$Q_{1,2,3}^{cu}$ by the replacements $A \leftrightarrow -A$ and $P
\leftrightarrow -P$.

We perform the matching at one-loop at the matching scale $\mu\gtrsim m_W$. Some
of the contributions generate logarithmic divergencies, which are canceled by the
appropriate counterterms, genuine short-distance contributions to the $|\Delta
c| =2$ Hamiltonian in Eq.~(\ref{eq:HdC=2}). We denote the corresponding
contributions to the $|\Delta c| =2$ Wilson coefficients $\delta
C^{cu(\prime)}_i$.  Using dimensional regularization with the $\overline{\rm
MS}$ prescription we obtain for the renormalized $|\Delta c| =2$ Wilson
coefficients
\begin{eqnarray}
C_1^{cu} &=& \delta C_1^{cu} + \frac{g^2}{32\pi^2}
  \sum_q \lambda_q\, (C^q_2-C^q_1)\, \ln \frac{\mu^2}{m_W^2}\,, \nonumber\\
C_4^{cu} &=& \delta C_4^{cu} - \frac{g^2}{16\pi^2}\sum_q
  \lambda_q\,  C^{q\prime}_{6}\, \ln \frac{\mu^2}{m_W^2} \,, \nonumber\\
C_5^{cu} &=& \delta C_5^{cu} - \frac{g^2}{16\pi^2}\sum_{q}
  \lambda_q\, C^{q\prime}_{5}\, \ln \frac{\mu^2}{m_W^2} \,,
\label{eq:dC2}
\end{eqnarray}
where here and below we neglect contributions proportional to $r_q=m_q^2/m_W^2$.
In particular, the leading order contributions to $C'_{1,2}$ and $C_{5,6}$ which
are proportional to $r_q \ln r_q$ were set to zero. Similarly, contributions of
the gluonic and electromagnetic dipole operators, $Q_{7,8}$, both at tree-level
via two insertions, as well as at one loop, are parametrically suppressed by
$r_c\, \alpha/\sin^2\theta_W$.\footnote{We have verified that due to similar
chiral suppression the contribution of $Q_{7(8)}$ to the down quark
(chromo)electric dipole moment via weak charged current ``dressing" remains well
below present bounds, even for order one Wilson coefficient $C_{7(8)}$.}
Numerically this leads to bounds of order unity on the corresponding Wilson
coefficients, well above the values obtained in Eq.~(\ref{eq:NP}), and thus no
useful constraint is obtained from $D-\bar D$ mixing.

To compute the contributions of $\mathcal H_{|\Delta c|=2}^{\rm eff}$ to
$M_{12}$, we take into account the running and mixing of the operators between
the matching scale $\mu$ and the scale $m_D$. This is performed using the
formula~\cite{Bona:2007vi}
\begin{eqnarray}
\bra{\bar D^0} \mathcal H_{|\Delta c|=2}^{\rm eff} \ket{D^0}_i 
  &=& \frac{G_F}{\sqrt 2} \sum_{j=1}^5 \sum_{r=1}^5 \left( b_j^{(r,i)}
  + \eta\, c_{j}^{(r,i)}\right) \eta^{a_j} \nonumber\\
&&{}\times C^{cu}_i(\mu) \bra{\bar D^0} Q_r^{cu} \ket{D^0}\,,
\end{eqnarray}
where all the relevant parameters are defined in Ref.~\cite{Bona:2007vi},
including the relevant hadronic operator matrix elements. Requiring that such
contributions do not exceed the bounds on $x_{12}$ and $x_{12}\sin\phi_{12}$ in
Eq.~(\ref{eq:expth}), we obtain the bounds on $C^{cu}_i$ at the matching scale
$\mu\sim1$\,TeV
\begin{align}
|C^{cu}_1| &\lesssim 5.7 \times 10^{-8}\,, & {\rm Im} (C^{cu}_1) \lesssim 1.6\times 10^{-8}\,, \nonumber\\
|C^{cu}_2| &\lesssim 1.6 \times 10^{-8}\,, & {\rm Im} (C^{cu}_2) \lesssim 4.3\times 10^{-9}\,, \nonumber\\
|C^{cu}_3| &\lesssim 5.8 \times 10^{-8}\,, & {\rm Im} (C^{cu}_3) \lesssim 1.6\times 10^{-8}\,, \nonumber\\
|C^{cu}_4| &\lesssim 5.6 \times 10^{-9}\,, & {\rm Im} (C^{cu}_4) \lesssim 1.6\times 10^{-9}\,, \nonumber\\
|C^{cu}_5| &\lesssim 1.6 \times 10^{-8}\,, & {\rm Im} (C^{cu}_5) \lesssim 4.5\times 10^{-9}\,. 
\end{align}

Inserting expressions (\ref{eq:dC2}) into the above constraints we can obtain
bounds on the combinations of $\delta C^{cu}_i$ and $C_i^q$ at the high scale.
In the following we put all counter term contributions to zero and consider only
a single chirality operator structure at a time. 

In order to control the QCD induced RGE evolution of the $|\Delta c|=1$
operators between the matching scale and the hadronic charm scale $\mu_D\sim
2$\,GeV, it is convenient to change flavor basis and consider the following set
of operators, both for $|\Delta c|=1$ (and $|\Delta s|=1$, see below) NP
Hamiltonians ($i=1,2,5,6$):
\begin{eqnarray}
Q^{(s-d)}_i  &=&  Q^{s}_i  - Q^{d}_i\,,  \nonumber \\
Q^{(c-u)}_i  &=&  Q^{c}_i  - Q^{u}_i\,,  \nonumber  \\ 
Q^{(8d)}_i   &=&   Q^{s}_i  + Q^{d}_i - 2 Q^{b}_i \,,   \nonumber  \\  
Q^{(b)}_i    &=&   Q^{s}_i  + Q^{d}_i +  Q^{b}_i - (3/2)\big(Q^{c}_i  + Q^{u}_i\big)\,,   \nonumber \\  
Q^{(0)}_i    &=&   Q^{s}_i  + Q^{d}_i +  Q^{b}_i + Q^{c}_i  + Q^{u}_i~,   \label{eq:newbasis}
\end{eqnarray}
and similarly for the primed operators.
With this choice, the $Q^{(0)(\prime)}_i$ are the standard QCD penguin operators, whose
RGE  evolution can  be found, for instance, in~\cite{Golden:1989qx}.   Moreover,
penguin contractions are completely absent in the RGE evolution at $\mu\gtrsim
m_c$ of the first two sets of terms in (\ref{eq:newbasis}) and, to a good
approximation (i.e., for $\mu\gtrsim m_b$), are safely negligible also in the
case of $Q^{(b,8d)(\prime)}_i$. For these operators we can thus consider, to
lowest order, a simplified RGE evolution in terms of  $2\times 2$ blocks of same
flavor and chirality:
\begin{eqnarray}
\frac{d C^{(f)}_i }{d\ln\mu} &=& \gamma_{LL}^T\, C^{(f)}_i , \quad 
\gamma^{(0)}_{LL} = \left( \begin{array}{cc} -\frac{6}{N_c} & 6 \\
  6 & -\frac{6}{N_c}  \end{array} \right) , \quad{i=1,2}\,, \nonumber \\
\frac{d C^{(f)}_i }{d\ln\mu} &=& \gamma_{LR}^T\, C^{(f)}_i , \quad 
\gamma^{(0)}_{LR} = \left( \begin{array}{cc}  \frac{6}{N_c}  & -6 \\
  0 & 6\frac{N_c^2-1}{N_c} \end{array} \right) , \quad{i=5,6}\,, \nonumber \\
\label{eq:rge}
\end{eqnarray}
where $f=\{s-d,\, c-u,\, 8d,\, b\}$, $N_c=3$ is the number of colors, and the
same equations hold for primed operators.

This basis also has the benefit of clearly distinguishing between various
contributions to $D-\bar D$ mixing observables suppressed by different CKM
prefactors. Most severe constraints are expected for the flavor combination
$Q_i^{(s-d)(\prime)}$ proportional to $\lambda_s-\lambda_d \approx 2
\lambda_s$. On the other hand, $Q_i^{(8d)(\prime)}$ contributions are
suppressed by $\lambda_s + \lambda_d - 2\lambda_b \approx - 3 \lambda_b$. An
even stronger suppression of $r_b \lambda_b$ is expected for the flavor
combinations $Q_i^{(b,0)(\prime)}$,  while 
$Q_i^{(c-u)(\prime)}$ do not contribute to $|\Delta c|=2$ observables at one
electroweak loop order. 

\begin{figure}[t]
  \centering
  \includegraphics[width=0.4\textwidth]{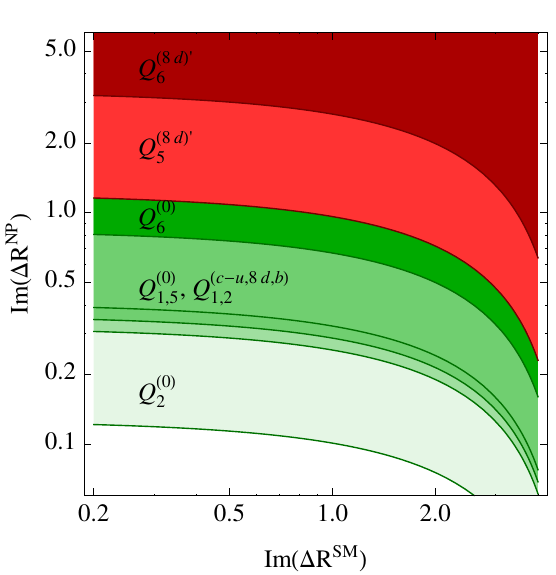}
  \caption{NP contributions of the form $Q^{(c-u,8d,b,0)}_{1,2}$,
  $Q^{(0)}_{5,6}$, and $Q_{5,6}^{(8d)\prime}$, reproducing the measured central
  value of $\Delta a_{CP}$ and consistent with searches for CP violation in
  $D-\bar D$ mixing (in the electronic version shaded in red) and the measured
  value of $\epsilon'/\epsilon$ (in the electronic version shaded in green).
  Both are estimated via one weak loop matching at $\mu\simeq 1$~TeV, as a
  function of the unknown amplitude ratios, $\Delta R^\SM$ and $\Delta
  R^\NP$ defined in Eq.~(\ref{eq:acpNP}). Assuming reasonable ranges for the
  hadronic matrix elements, contributions of individual operators can be consistent with all experimental
  results above the contours below the respective operator labels. 
  \label{fig:corr}}
\end{figure}

Considering thus only the cases $Q_i^{(s-d)(\prime)}$ and 
$Q_i^{(8d)(\prime)}$, we obtain the bounds on $C^q_i$ in
Table~\ref{tab:uvbounds}. We also verified that due to $r_q$ suppression,
$C'_{1,2}$, $C_{5,6}$, and $C_{7,8}$, as well as the contributions of
$C^{(b,0)}_{12}$ and $C^{(b,0)\prime}_{5,6}$ are presently allowed by $D -\bar
D$ data to be $\mathcal O(1)$.  We observe that $Q^{(s-d)}_{1,2}$ and
$Q^{(s-d)\prime}_{5,6}$ are excluded from explaining the central value of
$\Delta a_{CP}$ in Eq.~(\ref{eq:acpExp}) for $|\Delta R^{\SM}|\ll5$ and
reasonable values of $|\Delta R^{\NP}|$.  On the other hand,
$Q^{(8d)}_i$ can satisfy all present experimental constraints in the charm
sector given significant values of $|\Delta R^{\NP}|$ as also shown in
Fig.~\ref{fig:corr}.

\begin{table}[bt]
\tabcolsep 8pt
\begin{tabular}{c|cc}
\hline
$f$ & $s-d$ & $8d$  \\
\hline\hline
$\mathrm{Im}\big(C_{1,2}^{(f)}\big)$ & $5.4\times 10^{-6}$ & $4.5\times 10^{-3}$  \\
$\mathrm{Im}\big(C_5^{(f)\prime}\big)$ & $7.3\times 10^{-7}$ & $6.1\times 10^{-4}$  \\
$\mathrm{Im}\big(C_6^{(f)\prime}\big)$ & $2.7\times 10^{-7}$ & $2.2\times 10^{-4}$  \\
\hline
\end{tabular}
\caption{\label{tab:uvbounds} Bounds on the imaginary parts of $|\Delta c|=1$ Wilson
coefficients at the scale $\mu=1$~TeV, derived from searches for CP violation in
$D- \bar D$ mixing.}
\end{table}

\subsection{\boldmath Bounds on NP effects from $\epsilon'/\epsilon$}

As before, we can derive bounds from $T\{\mathcal H^{\rm
eff-NP}_{|\Delta c| = 1}(x)\, H^{\rm SM}_{c.c}(0)\}$ generating an effective
$|\Delta s| = 1$ interaction. We first project the  $|\Delta s| = 1$ effective
operators onto the following basis:
\be
\mathcal H^{\rm eff-\NP}_{|\Delta s| = 1} = \frac{G_F}{\sqrt 2} \sum_{i,q} C^{q(ds)}_i  Q_i^{q(ds)} + {\rm H.c.}\,,
\ee
where
\begin{eqnarray}
Q^{q(ds)}_1 &=& (\bar d s)_{V-A}\, (\bar q q)_{V-A}\,, \nonumber\\
Q^{q(ds)}_2 &=& (\bar d_\alpha  s_\beta)_{V-A}\, (\bar q_\beta q_\alpha)_{V-A}\,, \nonumber\\
Q^{q(ds)}_5 &=& (\bar d s)_{V-A}\, (\bar q q)_{V+A}\,, \nonumber\\  
Q^{q(ds)}_6 &=& (\bar d_\alpha s_\beta)_{V-A}\, (\bar q_\beta q_\alpha)_{V+A}\,.
\end{eqnarray}
These are the only effective operators generated at
the one-loop level from $T\{\mathcal H^{\rm eff-NP}_{|\Delta c| = 1}(x)\, H^{\rm
SM}_{c.c}(0)\}$ in the limit where we neglect light quark masses.  It is also
clear that these receive non-suppressed contributions only from the $Q_i^{q}$ in
$H^{\rm eff-NP}_{|\Delta c| = 1}$: the contributions of $Q^{q\prime}_{i}$ and
dipole operators are doubly Yukawa suppressed (in addition to the loop
suppression), and thus can be safely neglected. 

\begin{table}[t]
\tabcolsep 8pt
\begin{tabular}{c|ccccc}
\hline
$f$ & $s-d$ & $c-u$ & $8d$ & $b$ & $0$  \\
\hline\hline
${\rm Im}\big(C_{1}^{(f)}\big)\times10^{3}$ & $2.0$ & $2.0$ & $2.0$ & $0.79$ & $2.2$  \\
${\rm Im}\big(C_{2}^{(f)}\big)\times10^{3}$ & $2.0$ & $2.3$ & $2.0$ & $0.88$ & $6.6$  \\
${\rm Im}\big(C_{5}^{(f)}\big)\times10^{5}$ & $2.7$ & $2.8$ & $2.7$ & $1.1$ & $142$  \\
${\rm Im}\big(C_{6}^{(f)}\big)\times10^{5}$ & $0.90$ & $0.94$ & $0.90$ & $0.37$ & $28$  \\
\hline
\end{tabular}
\caption{\label{tab:ds} Bounds on the imaginary parts of $|\Delta c|=1$ Wilson coefficients at the scale $\mu =1$~TeV, from the contributions to $|\epsilon'/\epsilon|$.}
\end{table}

Procceding as before, we get
\begin{equation}
C^{q(ds)}_i = \delta C^{q(ds)}_i + C^q_{i} \frac{g^2}{32\pi^2}\, 
  \ln \frac{\mu^2}{m_W^2}\,, 
\label{eq:dsmatch}
\end{equation}
for all the relevant four-quark operators.
To compute the contributions of $\mathcal H^{\rm eff}_{|\Delta s| = 1}$  to
$K\to\pi\pi$ amplitudes we need to take into account the running and mixing of
the operators between the matching scale and a scale $\mu\sim 1$~GeV. Again it is
done in the flavor basis (\ref{eq:newbasis}), and using Eq.~(\ref{eq:rge})
analogous to the $|\Delta c|=1$ sector. 
The master formula for $\epsilon^\prime/\epsilon$ is 
\begin{eqnarray}
\left| \frac{\epsilon^\prime}{\epsilon} \right| 
  &=& \frac{\omega}{\sqrt{2}\, |\epsilon|\, {\rm Re}A_0} 
  \left| {\rm Im} A_0  -\frac{1}{\omega}\, {\rm Im} A_2 \right| , \\[4pt]
{\rm Im} A_I &=& \frac{G_F}{\sqrt{2}}\, \sum_{i,f} C^{(f)(ds)}_i\,
  \langle (2\pi)_I | Q_i^{(f)(ds)} | K \rangle\,,  \nonumber
\end{eqnarray}
where $\omega=  {\rm Re}A_2 / {\rm Re}A_0 \approx 0.045$ (from now on we omit the superscript
($sd$) on the coefficients and operators of the $|\Delta s|=1$ Hamiltonian).
Evaluating the matrix elements of $\mathcal H^{\rm eff-NP}_{|\Delta s| = 1}$  
in the large $N_c$ limit leads to 
\begin{eqnarray}
\left| \frac{\epsilon^\prime}{\epsilon} \right|_{\rm NP} &\approx&
  10^2\, \bigg| {\rm Im} \Big[ 
3.5  C_1^{(3/2)}   +3.4   C^{(3/2)}_2  -1.7\rho^2  C_5^{(3/2)}  \nonumber \\
&& -5.2 \rho^2 C_6^{(3/2)}  - 0.04 C_1^{(1/2)} - 0.12  C_2^{(1/2)} 
\nonumber \\
&&   - 0.04\rho^2 C_5^{(1/2)}  + 0.11\rho^2 C_6^{(1/2)}  \Big] \bigg|\,,
\end{eqnarray}
in terms of the $|\Delta s|=1$ Wilson coefficients at the low scale $(\mu= 1.4\,\mathrm{GeV})$, where $C_i^{(3/2)}=
[-C^{(s-d)}+C^{(c-u)}+C^{(8d)}]/2+(5/4)C^{(b)}$,
$C_i^{(1/2)}=[C^{(s-d)}+C^{(c-u)}-C^{(8d)}]/2+(1/4)C^{(b)}-C^{(0)}$, 
and $\rho=m_K/m_s$.  Imposing  the conservative bound $|
\epsilon^\prime/ \epsilon |_{\rm NP} <  | \epsilon^\prime/\epsilon |_{\rm exp}
\approx 1.7 \times 10^{-3}$,  leads to severe constraints on all the
coefficients. In terms of $|\Delta s| =1$ Wilson coefficients at the high scale
($\mu=1$~TeV) the constraints read
\begin{align}
\mathrm{Im} (C^{(s-d)}_{1}) &\lesssim 1.4 \times 10^{-5}\,, & \mathrm{Im} (C^{(s-d)}_{2}) &\lesssim 1.4 \times 10^{-5}\,, \nonumber\\
\mathrm{Im} (C^{(s-d)}_{5}) &\lesssim 1.9 \times 10^{-7}\,, & \mathrm{Im} (C^{(s-d)}_{6}) &\lesssim 6.1 \times 10^{-8}\,, \nonumber\\
\mathrm{Im} (C^{(c-u)}_{1}) &\lesssim 1.3 \times 10^{-5}\,, & \mathrm{Im} (C^{(c-u)}_{2}) &\lesssim 1.6 \times 10^{-5}\,, \nonumber\\
\mathrm{Im} (C^{(c-u)}_{5}) &\lesssim 1.9 \times 10^{-7}\,, & \mathrm{Im} (C^{(c-u)}_{6}) &\lesssim 6.4 \times 10^{-8}\,, \nonumber\\
\mathrm{Im} (C^{(8d)}_{1}) &\lesssim 1.4 \times 10^{-5}\,, & \mathrm{Im} (C^{(8d)}_{2}) &\lesssim 1.4 \times 10^{-5}\,, \nonumber\\
\mathrm{Im} (C^{(8d)}_{5}) &\lesssim 1.9 \times 10^{-7}\,, & \mathrm{Im} (C^{(8d)}_{6}) &\lesssim 6.1 \times 10^{-8}\,, \nonumber\\
\mathrm{Im} (C^{(b)}_{1}) &\lesssim 5.4 \times 10^{-6} \,,& \mathrm{Im} (C^{(b)}_{2}) &\lesssim 5.9 \times 10^{-6}\,,\nonumber\\
\mathrm{Im} (C^{(b)}_{5}) &\lesssim 7.5 \times 10^{-8} \,,& \mathrm{Im} (C^{(b)}_{6}) &\lesssim 2.5 \times 10^{-8}\,,\nonumber\\
\mathrm{Im} (C^{(0)}_{1}) &\lesssim 1.5 \times 10^{-5} \,,& \mathrm{Im} (C^{(0)}_{2}) &\lesssim 4.5 \times 10^{-5}\,,\nonumber\\
\mathrm{Im} (C^{(0)}_{5}) &\lesssim 9.6 \times 10^{-6} \,,& \mathrm{Im} (C^{(0)}_{6}) &\lesssim 1.9 \times 10^{-6}\,.
\end{align}
Inserting the matching conditions (\ref{eq:dsmatch}), we obtain bounds on the
$|\Delta c|=1$ Wilson coefficients in Table~\ref{tab:ds}. We observe that all
$Q_{5,6}^{(f)}$ except $Q_{5,6}^{(0)}$ are excluded from contributing
significantly to $\Delta a_{CP}$. The remaining operators are only marginally
constrained and can give observable effects in the charm sector provided
$|\Delta R^{\NP}|$ have significant values as also shown in Fig.~\ref{fig:corr}.

\begin{table}[t]
\tabcolsep 8pt
\begin{tabular}{ccc}
\hline
Allowed & Ajar & Disfavored \\
\hline\hline
$Q_{7,8}\,,\ Q'_{7,8}\,, $ 
  & $ Q_{1,2}^{(c-u,8d,b,0)},$
  & $Q_{1,2}^{s-d}\,, \ Q_{5,6}^{(s-d)\prime},$\\
$ {\forall f}\ Q_{1,2}^{f\prime}\,,\ Q_{5,6}^{(c-u,b,0)\prime} $ 
  & $Q_{5,6}^{(0)}\,,\ Q_{5,6}^{(8d)\prime}$
  & $ Q_{5,6}^{s-d,c-u,8d,b}$\\
\hline
\end{tabular}
\caption{\label{tab:summary} List of $|\Delta c|=1$ operators grouped according to whether they
can contribute to $\Delta a_{CP}$ at a level comparable to the central value of
the measurement, given the constrains from $D-\bar D$ mixing and
$\epsilon'/\epsilon$.}
\end{table}

\section{Conclusions}

We explored the implications of the recent LHCb measurement of a $3.5\sigma$
deviation from no CP violation in $D$ decays.  Clearly, it will require more
data to establish whether the measurement is or is not consistent with the SM. 
While a sufficient QCD enhancement of the penguin matrix element cannot be
excluded at the present time, if similar CP violation is observed in other
channels as well (e.g., pseudoscalar-vector final states, three-body decays,
$D_s$ or $\Lambda_c$ decays), then it would suggest that the measurement is due
to new short distance physics, rather than the enhancement of a hadronic matrix
element in one particular channel.

Our analysis implies that operators where the charm
bilinear current is of $V-A$ structure are constrained by $D-\bar D$
mixing or by $\epsilon'/\epsilon$, especially the ones which violate $U$-spin. A complete list of the operators grouped
according to whether they can contribute to $\Delta a_{CP}$ at a level
comparable to the central  value of the measurement, given the constrains from
$D-\bar D$ mixing and $\epsilon'/\epsilon$, is shown in
Table~\ref{tab:summary}.   It is also worth noting that in cases where the new
physics contributions are large, we generically expect sizable contributions to
CP violation in $D-\bar D$ mixing (and in $\epsilon'/\epsilon$) to arise. This
will be tested when the constraints on CP violation in $D-\bar D$ mixing will
improve substantially with more LHCb and future super-$B$-factory data.

\begin{acknowledgments}

We thank Marco Gersabeck, Vladimir Gligorov, and Alex Kagan for
helpful discussions.
GI acknowledges the support  of the TU~M\"unchen -- Institute for Advanced
Study, funded by the German Excellence Initiative, and the  EU ERC Advanced
Grant FLAVOUR (267104).
The work of JFK was supported in part  by the Slovenian Research Agency. 
The work of ZL was supported in part by the Director, Office of Science, Office
of High Energy Physics of the U.S.\ Department of Energy under contract
DE-AC02-05CH11231. GP is supported by the GIF, Gruber foundation, IRG, ISF and Minerva.

\end{acknowledgments}

\end{document}